\documentclass[%
preprint, 
superscriptaddress,
amsmath,amssymb,
aps, 
prb,
]{revtex4-2}
\makeatletter
\let\jpsjoriginalbibitem\bibitem
\renewcommand{\bibitem}{%
  \@ifnextchar[{\jpsjbibitem}{\jpsjoriginalbibitem}}
\def\jpsjbibitem[#1]#2{%
  \jpsjoriginalbibitem{#2}}
\makeatother
\usepackage{graphicx}
\usepackage{dcolumn}
\usepackage{bm}
\usepackage{color}
\usepackage[dvipsnames]{xcolor}
\usepackage[draft]{hyperref}
\begin{document}
\title{Two Microscopic Mechanisms of Piezomagnetism in CoF$_2$\\ from First-Principles Calculations}

\author{Hiroshi Katsumoto}
\email{Contact author: hiroshi.katsumoto@phys.s.u-tokyo.ac.jp}
\altaffiliation[Present address: ]%
{Department of Physics, The University of Tokyo, Japan}
 \affiliation{Division of Materials and Manufacturing Science, Graduate School of Engineering, The University of Osaka, Suita, Osaka 565-0871, Japan}
 \affiliation{Center for Spintronics Research Network, The University of Osaka, Toyonaka, Osaka 560-8531, Japan}

\author{Kunihiko Yamauchi}%
 \affiliation{Center for Spintronics Research Network, The University of Osaka, Toyonaka, Osaka 560-8531, Japan}
 \affiliation{Department of Precision Engineering, Graduate School of Engineering, The University of Osaka, Suita, Osaka 565-0871, Japan}%

\author{Tamio Oguchi}
 \affiliation{Center for Spintronics Research Network, The University of Osaka, Toyonaka, Osaka 560-8531, Japan}
 \affiliation{Department of Precision Engineering, Graduate School of Engineering, The University of Osaka, Suita, Osaka 565-0871, Japan}%
 
\date{\today}

\begin{abstract}
    Rutile-structured CoF$_2$ has long been recognized as a prototypical piezomagnetic material. Recently, it has attracted renewed interest as an altermagnet, exhibiting spin-split electronic bands even in the absence of spin-orbit coupling. Although the piezomagnetic response of CoF$_2$ has been extensively discussed from the viewpoint of magnetic symmetry, its microscopic origin has remained elusive.
First-principles calculations reveal two distinct microscopic mechanisms of piezomagnetism in CoF$_2$. Under $xy$ shear strain, the local volumes of the CoF$_6$ octahedra surrounding the two Co sites become different, leading to unequal magnetic moments on the two sublattices and hence a net magnetization. In contrast, under $yz$ shear strain, the piezomagnetic response originates from spin canting induced by the Dzyaloshinskii--Moriya interaction through spin-orbit coupling. The presence of two distinct microscopic mechanisms may be a general feature of piezomagnetic antiferromagnets. 
\end{abstract}

\maketitle


\section{Introduction}
The piezomagnetic effect, namely the linear induction of magnetization by applied mechanical strain or stress, is a fundamental manifestation of spin--lattice coupling in antiferromagnets. The piezomagnetic effect in rutile antiferromagnets was first predicted from magnetic symmetry by Dzialoshinskii~\cite{Dzialoshinskii1958-xq}. It was subsequently observed experimentally in MnF$_2$ and CoF$_2$~\cite{Borovik-Romanov1960-mi,Baruchel1988-pl}. Moriya later proposed a microscopic mechanism for the piezomagnetic response of rutile fluorides MnF$_2$, FeF$_2$, and CoF$_2$, showing that spin--orbit coupling (SOC), through the Dzyaloshinskii--Moriya interaction (DMI) and magnetic anisotropy, plays an essential role in generating piezomagnetism~\cite{Dzialoshinskii1957-vg,Moriya1960-pu,Moriya1959-gy}. These pioneering studies established the symmetry conditions and microscopic understanding of piezomagnetism in antiferromagnets.

The recent discovery of altermagnetism has renewed interest in piezomagnetism~\cite{Smejkal2022-cm,Jungwirth2026-rn,Thao2023CaCrO3, Osumi2024MnTe, Osumi2026RuO2}. Altermagnets exhibit spin-split electronic bands even in collinear antiferromagnets and in the absence of SOC~\cite{Noda2016-wh,Okugawa2018-is,Naka2019-px}, making them promising candidates for next-generation spintronic devices~\cite{Takagi2025-sl,Jungwirth2026-pd}. Their spin splitting originates from crystal and magnetic symmetries rather than relativistic interactions, providing a new perspective on magnetically driven electronic phenomena.

Motivated by these developments, the relationship between altermagnetism and piezomagnetism has recently attracted considerable attention. Aoyama and Ohgushi experimentally demonstrated a sizable piezomagnetic effect in altermagnetic MnTe and showed that piezomagnetism provides a sensitive probe of global time-reversal-symmetry breaking~\cite{AoyamaPRM2024}. They further investigated the microscopic origin of the piezomagnetic response by combining symmetry analysis with first-principles calculations~\cite{Aoyama2024-tm,Komuro2025-ot,Nanjo2025-lk}. On the theoretical side, recent studies have shown that nonrelativistic piezomagnetism can emerge under appropriate electronic conditions, including in hole-doped systems and organic altermagnets~\cite{Ma2021-bi,Streltsov2025-ev,Naka2025-bt}. These studies suggest that SOC is not always indispensable for piezomagnetism, but its microscopic origin remains far from fully understood.

CoF$_2$ provides an ideal platform to address this issue because it is both a prototypical piezomagnetic material and a representative altermagnet. While its piezomagnetic effect has historically been interpreted within the framework of SOC-based mechanisms, it remains unclear whether the spin-split electronic structure characteristic of altermagnets can itself generate piezomagnetism. In this work, we perform first-principles electronic-structure calculations for CoF$_2$ and systematically investigate its piezomagnetic response under different shear strains with and without SOC.

\section{Crystal Structure and Magnetic Order}

\begin{figure}[t]
\begin{center}
    \includegraphics[width=0.45\textwidth]{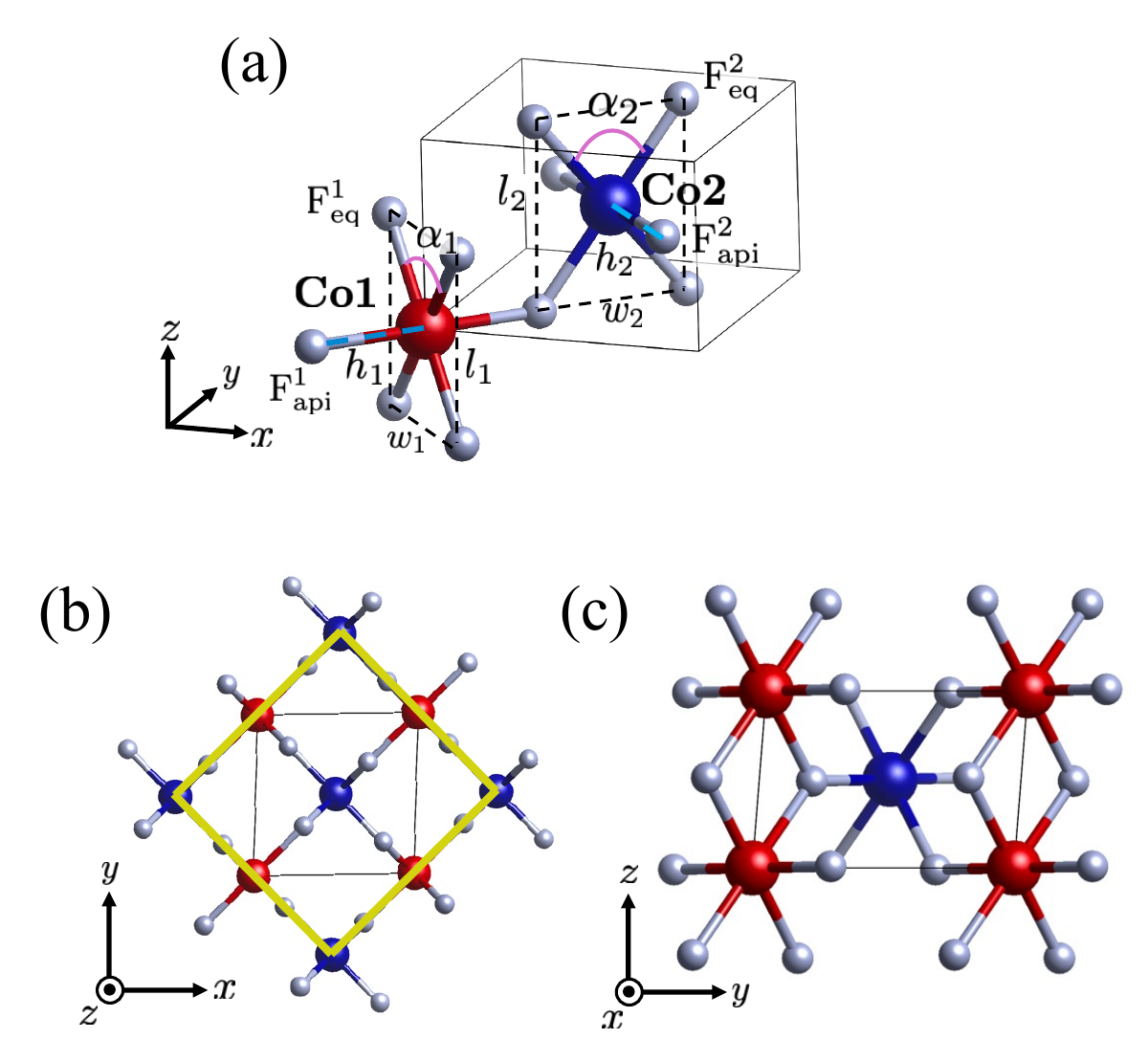}
\end{center}
\caption{
Crystal structures of CoF$_2$. In this paper, the up-spin and down-spin Co sites are referred to as Co1 and Co2, respectively. They are shown as red and blue spheres, while fluorine atoms are shown as silver spheres. The magnetic moments are aligned antiparallel along the $z$ axis. 
(a) Undistorted rutile structure with space group $P4_{2}/mnm$. The CoF$_6$ octahedron consists of two apical and four equatorial Co--F bonds. The four equatorial Co--F bonds form a rectangle with side lengths $w_{1/2}$ and $l_{1/2}$ (black dashed lines), while the apical Co--F bond length is denoted by $h_{1/2}$ (blue dashed line). The angle between two adjacent equatorial Co--F bonds is denoted by $\alpha_{1/2}$.
(b) Crystal structure under $xy$ shear strain. The symmetry is reduced to $Cmmm$, making the Co1 and Co2 sites crystallographically inequivalent, with Wyckoff positions $2c$ and $2a$, respectively. The yellow rectangle indicates the orthorhombic unit cell in the $xy$ plane.
(c) Crystal structure under $yz$ shear strain. The symmetry is reduced to $P2_1/c$, while the Co1 and Co2 sites remain crystallographically equivalent, both occupying the Wyckoff position $2c$.
}
\label{fig:all-str}
\end{figure}

CoF$_2$ crystallizes in the rutile structure, as do MnF$_2$ and FeF$_2$, which constitute a family of antiferromagnetic rutile fluorides. As shown in Fig.~\ref{fig:all-str}(a), the crystal structure belongs to the space group $P4_2/mnm$. The unit cell contains two crystallographically equivalent Co sites. Throughout this paper, the up-spin and down-spin Co sites are referred to as Co1 and Co2, respectively, in the antiferromagnetic ground state. The magnetic moments are aligned antiparallel along the crystallographic $z$ axis when SOC is considered. Each Co atom is surrounded by a slightly distorted CoF$_6$ octahedron consisting of two apical and four equatorial Co--F bonds. The four equatorial fluorine atoms form a rectangle with side lengths $w$ and $l$, while the distance between the Co atom and the apical fluorine atom is denoted by $h$. The F$_{\rm eq}$--Co--F$_{\rm eq}$ bond angle is denoted by $\alpha$. 

The equilibrium magnetic structure is described by the magnetic space group $P4_{2}^{\prime}/mnm^{\prime}$ with the magnetic point group $4^{\prime}/mmm^{\prime}$. According to the symmetry, only two independent piezomagnetic tensor components are allowed: $Q_{14}(=Q_{25})$ and $Q_{36}$. Consequently, an $xy$ shear strain induces a finite magnetization along the $z$ direction, whereas a $yz$ shear strain induces a finite magnetization along the $x$ direction.\cite{Komuro2025-ot}

\section{Computational Methods}

First-principles calculations were performed within density functional theory using the projector augmented-wave (PAW) method as implemented in the Vienna \textit{Ab initio} Simulation Package (VASP).\cite{Kresse1996,Kresse1999} The exchange--correlation functional was treated within the generalized gradient approximation in the Perdew--Burke--Ernzerhof (PBE) form with the Hubbard correction (GGA+$U$), where the effective on-site Coulomb interaction was set to $U$ = 2 eV\cite{Perdew1996,Dudarev1998}. Spin-polarized calculations were performed using the noncollinear formalism. Starting from the collinear antiferromagnetic spin configuration, both the directions and magnitudes of the local magnetic moments were determined self-consistently. 
The plane-wave cutoff energy was set to 500~eV. The Brillouin zone was sampled using a $\Gamma$-centered $10 \times 10 \times 12$ $k$-point mesh. The atomic positions were relaxed until the residual Hellmann--Feynman forces on each atom became smaller than $10^{-5}$~eV/\AA.   
Carrier doping was simulated by changing the total number of electrons. Charge neutrality was maintained by the uniform compensating background charge implemented in VASP.  Spin--orbit coupling (SOC) was included where explicitly stated.

The lattice parameters and internal atomic coordinates of the unstrained structure were first fully optimized. 
The strained lattice vectors were generated from the optimized equilibrium lattice vectors according to
\begin{equation}
\boldsymbol{a}_{i}^{\,\prime}
=
\left(\boldsymbol{I}+\boldsymbol{\varepsilon}\right)
\boldsymbol{a}_{i},
\end{equation}
where $\boldsymbol{I}$ is the identity matrix and
$\boldsymbol{\varepsilon}$ is the symmetric strain tensor.
The $xy$ and $yz$ shear strains were introduced using
\begin{equation}
\boldsymbol{\varepsilon}^{(xy)}
=
\begin{pmatrix}
0 & \varepsilon_{xy} & 0 \\
\varepsilon_{xy} & 0 & 0 \\
0 & 0 & 0
\end{pmatrix},
\qquad
\boldsymbol{\varepsilon}^{(yz)}
=
\begin{pmatrix}
0 & 0 & 0 \\
0 & 0 & \varepsilon_{yz} \\
0 & \varepsilon_{yz} & 0
\end{pmatrix}.
\end{equation}
The symmetric $xy$ shear strain corresponds to tensile and compressive strains along the $[110]$ and $[1\bar{1}0]$ directions, respectively, and lowers the crystal symmetry from tetragonal to orthorhombic. 
After applying each strain, the strained lattice vectors were fixed and only the internal atomic coordinates were relaxed. 
The strain-form piezomagnetic coefficient was evaluated from the slope of the magnetic moment per unit cell with respect to the applied strain,
\[
e^{\rm m}_{ij}
=
\frac{\partial M_i}{\partial \varepsilon_j},
\]
where $m_i$ denotes the magnetic moment per unit cell.
Figures~\ref{fig:all-str}(b) and \ref{fig:all-str}(c) show the crystal structures under $xy$- and $yz$-shear strains, respectively. Under $xy$ shear strain, the crystal symmetry is lowered to $Cmmm$, and the crystallographic symmetry operation exchanging the Co1 and Co2 sites is lost. Consequently, the two Co sites become crystallographically inequivalent, leading to different local geometries of the two CoF$_6$ octahedra. In contrast, under $yz$ shear strain, the crystal symmetry is lowered to $P2_{1}/c$, while the crystallographic symmetry operation exchanging the Co1 and Co2 sites is preserved. Consequently, the two Co sites remain crystallographically equivalent despite the reduced crystal symmetry.

\section{Results and Discussion}

\subsection{Piezomagnetism under $xy$ shear strain}

\begin{figure}[htbp]
\begin{center}
    \includegraphics[width=0.4\textwidth]{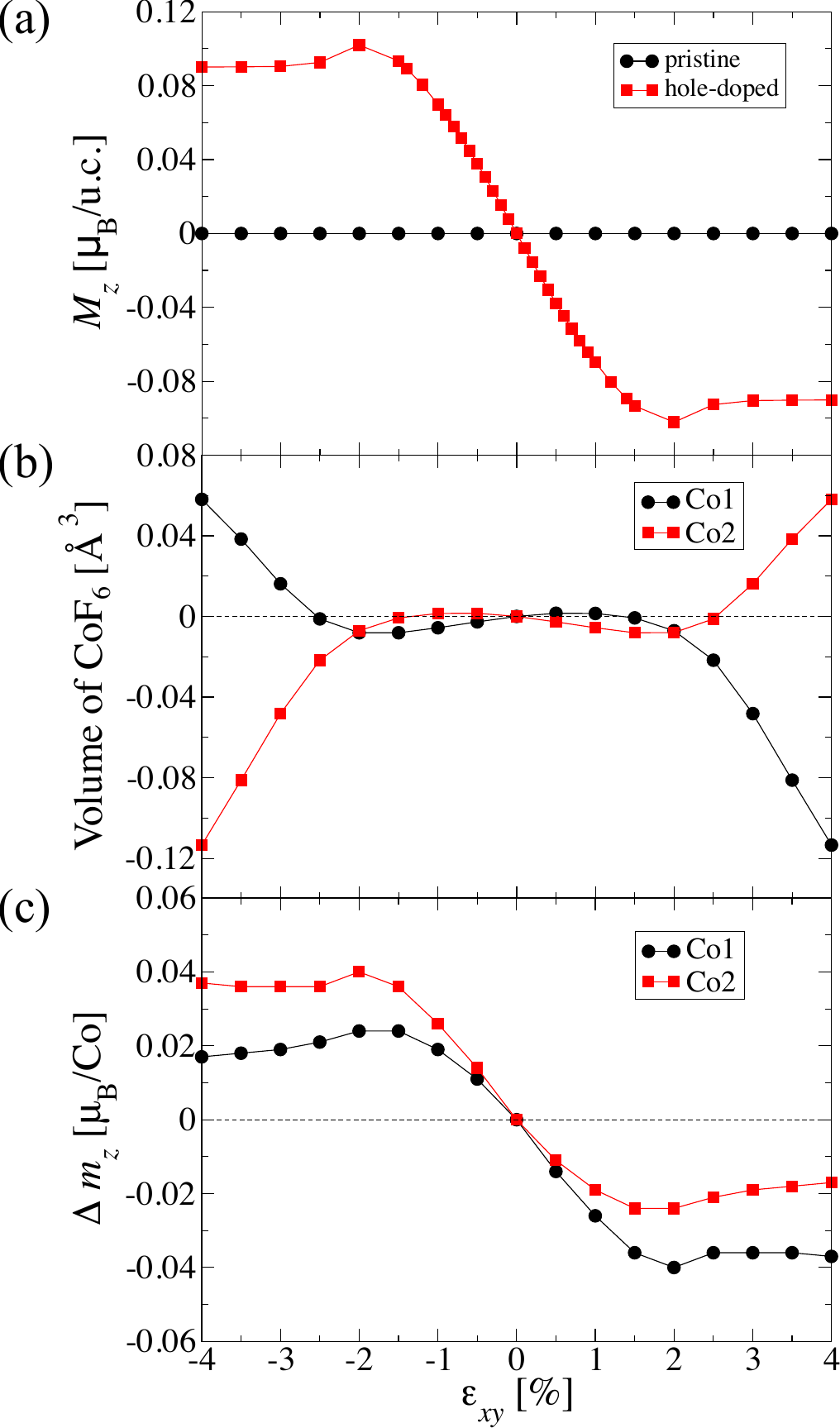}
    \caption{(a) Piezomagnetically induced magnetization along the $z$ axis as a function of the $xy$ shear strain without spin--orbit coupling. Circles and squares represent pristine and 0.1\% hole-doped CoF$_2$, respectively.
    (b) Changes in the CoF$_6$ octahedral volume (upper panel) and (c) the local magnetic moment along the $z$ axis (lower panel) for the Co1 (black circles) and Co2 (red squares) sites under the $xy$ shear strain.}
    \label{fig:piezomag_xy}
\end{center}
\end{figure}

Figure~\ref{fig:piezomag_xy}(a) shows the induced magnetization under the $xy$ shear strain without spin--orbit coupling (SOC). In pristine CoF$_2$, the induced magnetization is negligibly small. To clarify the microscopic mechanism, we also performed calculations for 0.1\% hole-doped CoF$_2$, which exhibits a substantially enhanced piezomagnetic response while preserving the same strain dependence. The induced magnetization is nearly proportional to strain for $|\varepsilon_{xy}| \leq 0.5$, corresponding to a piezomagnetic coefficient of $e^{\rm m}_{36}=-3.80~\mu_{\mathrm{B}}$/u.c. At larger strains, the response becomes nonlinear and gradually approaches saturation for $|\varepsilon_{xy}| \gtrsim 2.0$. The corresponding stress-form piezomagnetic coefficients are presented in the Supplemental Material.

Figures~\ref{fig:piezomag_xy}(b) and (c) show the strain dependence of the CoF$_6$ octahedral volume and the local magnetic moment at the Co1 and Co2 sites, respectively. Because the $xy$ shear strain removes the crystallographic symmetry relating the two Co sites, the two CoF$_6$ octahedra become inequivalent. Their volumes change in opposite directions with increasing strain, accompanied by corresponding changes in the local magnetic moments. As a result, the cancellation between the two antiferromagnetic sublattices becomes incomplete, giving rise to a finite net magnetization.

\subsection{Piezomagnetism under $yz$ shear strain}

\begin{figure}[htbp]
\begin{center}
    \includegraphics[width=0.4\textwidth]{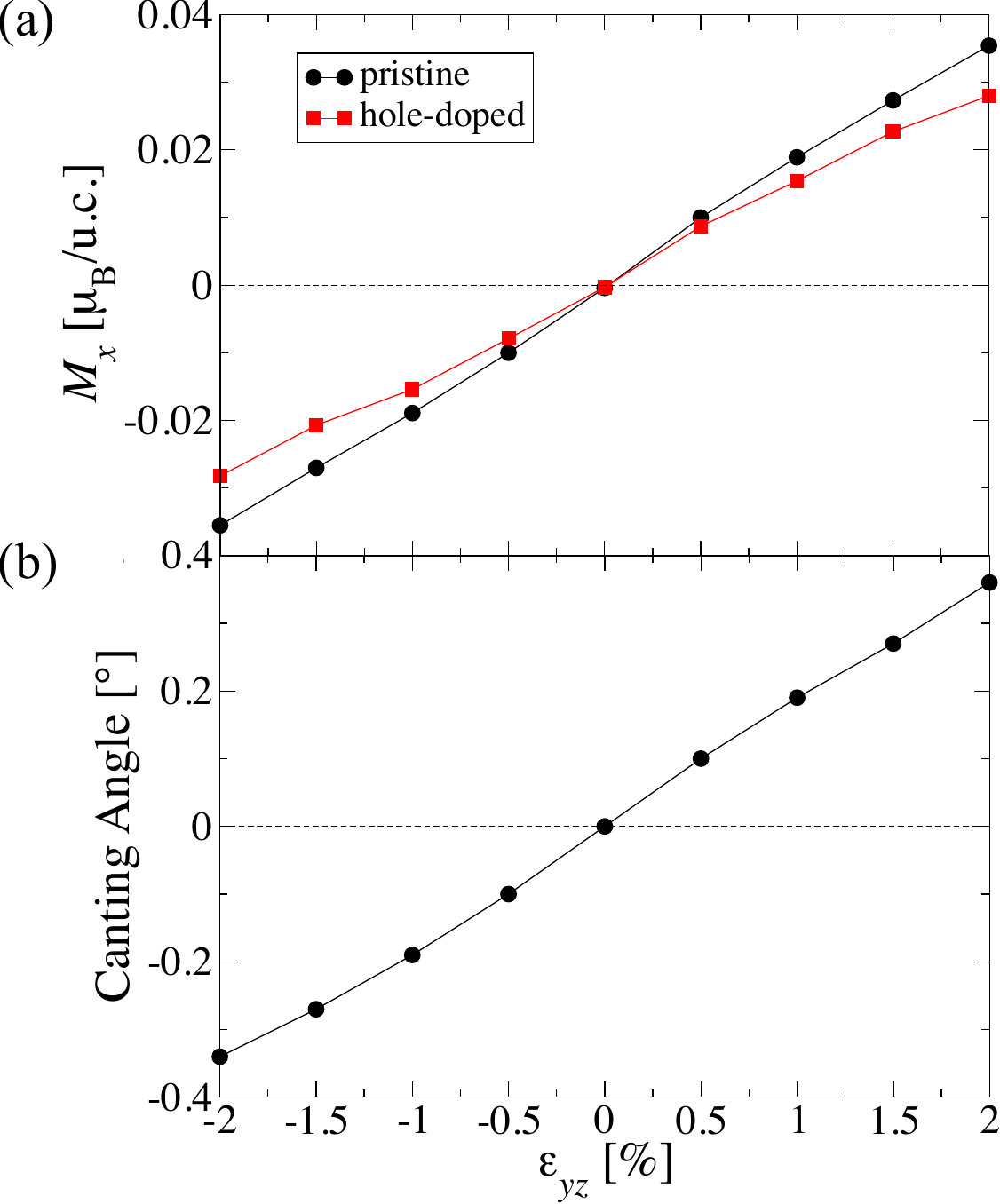}
    \caption{(a) Induced magnetization along the $x$ axis as a function of the $yz$ shear strain with spin--orbit coupling. Circles and squares represent pristine and hole-doped CoF$_2$, respectively. (b) Canting angle of the magnetic moments under the $yz$ shear strain in pristine CoF$_2$.}
    \label{fig:piezomag_yz}
\end{center}
\end{figure}

Figure~\ref{fig:piezomag_yz}(a) shows the induced magnetization under the $yz$ shear strain in the presence of spin--orbit coupling (SOC). In contrast to the $xy$ shear strain, the induced magnetization appears along the $x$ direction and increases almost linearly with the applied strain. A linear fit for $|\varepsilon_{yz}| \leq 1.0$ yields a piezomagnetic coefficient of $e^{\rm m}_{14}=0.72~\mu_{\mathrm{B}}$/u.c. for pristine CoF$_2$. The corresponding stress-form piezomagnetic coefficients are presented in the Supplemental Material. Hole doping has little effect on either the magnitude or the strain dependence of the induced magnetization, indicating that this piezomagnetic response is essentially insensitive to the carrier concentration.

Figure~\ref{fig:piezomag_yz}(b) shows the canting angle of the magnetic moments as a function of the $yz$ shear strain. In the absence of strain, the magnetic moments are collinear and aligned along the crystallographic $z$ axis. As the $yz$ shear strain increases, the canting angle increases continuously, demonstrating that the induced magnetization originates from strain-induced spin canting. This behavior is consistent with the Dzyaloshinskii--Moriya mechanism expected from the symmetry of the strained crystal.

\subsection{Electronic structure and microscopic mechanisms}

\begin{figure}[htb]
\begin{center}
    \includegraphics[width=0.4\textwidth]{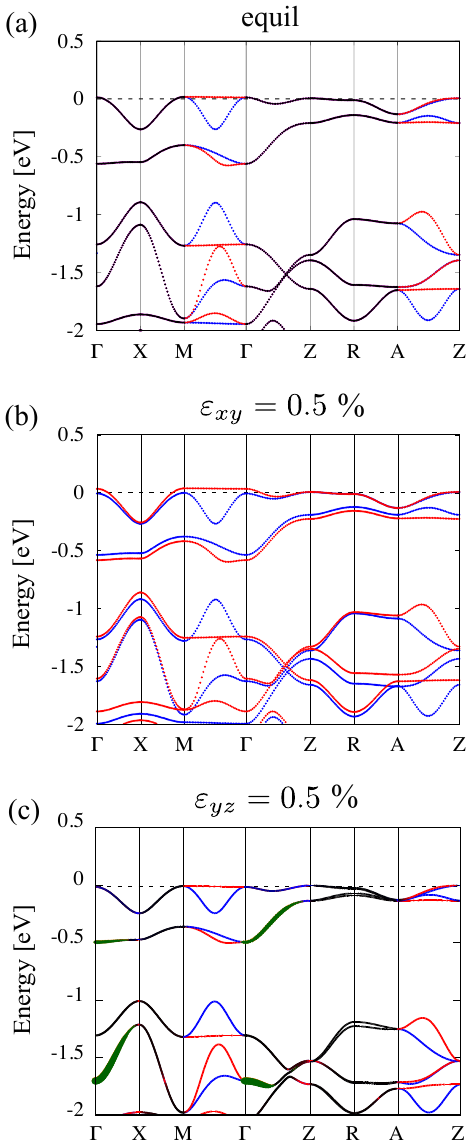}
\caption{
Electronic band structures of CoF$_2$. (a) Equilibrium structure and (b) structure under the $xy$ shear strain of $\varepsilon_{xy}=0.5\%$ for 0.1\% hole-doped CoF$_2$ without spin--orbit coupling (SOC). The red and blue dots represent the spin-up and spin-down bands, respectively. (c) Structure under the $yz$ shear strain of $\varepsilon_{yz}=0.5\%$ for undoped CoF$_2$ with SOC ($\mathbf{S}\parallel z$). The red and blue lines represent the $S_z$ component, while the green dots indicate the $S_x$ component (the maximum magnitude of the plotted $S_x$ component is approximately $+0.28$). Although the applied shear strains slightly lower the crystal symmetry from the tetragonal structure to orthorhombic or monoclinic structures, the corresponding changes in the Brillouin zone are neglected, and the high-symmetry points are labeled using those of the parent tetragonal Brillouin zone. The energy is measured from the Fermi level.
}
    \label{fig:bands}
\end{center}
\end{figure}

Figure~\ref{fig:bands} compares the electronic band structures of the equilibrium and strained structures. In the equilibrium structure without SOC, the magnetic space group breaks the combined space-inversion and time-reversal ($PT$) symmetry, giving rise to the characteristic altermagnetic spin splitting along the M--$\Gamma$ and A--Z directions. In contrast, the bands remain spin degenerate at the $\Gamma$ point.

Under the $xy$ shear strain, the crystallographic inequivalence of the two Co sites lifts the remaining spin degeneracy throughout the Brillouin zone, resulting in ferrimagnetism. As discussed in the Supplemental Material, the strain-induced band splitting modifies the occupations of the spin-polarized states near the Fermi level, leading to different local magnetic moments on the two Co sites. 

In contrast, the $yz$ shear strain produces only minor changes in the electronic band dispersion, and no pronounced spin splitting appears at the $\Gamma$ point. Instead, the SOC-induced spin canting generates a finite $S_x$ component around the $\Gamma$ point along the $\Gamma$--X and $\Gamma$--Z directions, indicating that the primary effect of the $yz$ shear strain is to modify the spin texture rather than the band dispersion itself.

These results demonstrate that CoF$_2$ possesses two distinct microscopic mechanisms of piezomagnetism. The $xy$ shear strain induces a nonrelativistic piezomagnetic response through crystallographically inequivalent Co sites and the resulting redistribution of electronic occupation. In contrast, the $yz$ shear strain produces a relativistic piezomagnetic response through SOC-induced spin canting associated with the Dzyaloshinskii--Moriya interaction. The enhancement of the $xy$ response by hole doping further highlights the importance of carrier redistribution between the two inequivalent Co sites.

Recent studies have pointed to a close relationship between altermagnetism and piezomagnetism.\cite{AoyamaPRM2024,Komuro2025-ot,Nanjo2025-lk} The present results suggest that both nonrelativistic and relativistic piezomagnetic mechanisms may be common among piezomagnetic altermagnets. Experimental measurements using single crystals would provide a direct test of these microscopic mechanisms. 

\section{Summary}

We have investigated the microscopic origin of piezomagnetism in rutile CoF$_2$ by first-principles calculations. Under the $xy$ shear strain, a finite magnetization is induced even without spin--orbit coupling through the inequivalence of the two Co sites, whereas the piezomagnetic response under the $yz$ shear strain originates from spin canting induced by spin--orbit coupling.
These results demonstrate that CoF$_2$ possesses two distinct microscopic mechanisms of piezomagnetism. The coexistence of nonrelativistic and relativistic piezomagnetic mechanisms may be a general feature of piezomagnetic antiferromagnets and motivates further studies on other candidate altermagnets.

\begin{acknowledgments}

The authors acknowledge valuable discussions with T. Aoyama, K. Kimura, and K. Sato. 
This work was supported  by JST-CREST (Grant No. JPMJCR22O2) and by
Institute for Open and Transdisciplinary Research Initiatives (OTRI), the University of Osaka.
The computation in this work has been done using the facilities of the Supercomputer Center, the Institute for Solid State Physics, the University of Tokyo. The crystallographic figure was generated using the VESTA program~\cite{Momma2011}. 
\end{acknowledgments}

\bibliographystyle{apsrev4-2}
\bibliography{ref}

@ARTICLE{Naka2025-bt,
  title     = "Nonrelativistic piezomagnetic effect in an organic altermagnet",
  author    = "Naka, Makoto and Motome, Yukitoshi and Miyazaki, Tsuyoshi and
               Seo, Hitoshi",
  journal   = "J. Phys. Soc. Jpn.",
  publisher = "Physical Society of Japan",
  volume    =  94,
  number    =  8,
  month     =  "15~" # aug,
  pages     =  083702,
  year      =  2025
}

@ARTICLE{Komuro2025-ot,
  title     = "Revisiting the piezomagnetic effect in the rutile fluorides
               {MnF2} and {CoF2}",
  author    = "Komuro, Minato and Aoyama, Takuya and Ohgushi, Kenya",
  journal   = "Phys. Rev. B.",
  publisher = "American Physical Society (APS)",
  volume    =  111,
  number    =  21,
  pages     =  214445,
  month     =  "30~" # jun,
  year      =  2025
}

@ARTICLE{Aoyama2024-tm,
  title     = "Piezomagnetic properties in altermagnetic {MnTe}",
  author    = "Aoyama, Takuya and Ohgushi, Kenya",
  journal   = "Phys. Rev. Mater.",
  publisher = "American Physical Society (APS)",
  volume    =  8,
  number    =  4,
  pages     = "L041402",
  month     =  "1~" # apr,
  year      =  2024
}

@ARTICLE{Smejkal2022-cm,
  title     = "Emerging research landscape of altermagnetism",
  author    = "Šmejkal, Libor and Sinova, Jairo and Jungwirth, Tomas",
  journal   = "Phys. Rev. X.",
  publisher = "American Physical Society (APS)",
  volume    =  12,
  number    =  4,
  pages     =  040501,
  month     =  "8~" # dec,
  year      =  2022
}

@ARTICLE{Ma2021-bi,
  title     = "Multifunctional antiferromagnetic materials with giant
               piezomagnetism and noncollinear spin current",
  author    = "Ma, Hai-Yang and Hu, Mengli and Li, Nana and Liu, Jianpeng and
               Yao, Wang and Jia, Jin-Feng and Liu, Junwei",
  journal   = "Nat. Commun.",
  publisher = "Springer Science and Business Media LLC",
  volume    =  12,
  number    =  1,
  pages     =  2846,
  month     =  "14~" # may,
  year      =  2021
}

@ARTICLE{Borovik-Romanov1960-mi,
  title    = "Piezomagnetism in the antiferromagnetic fluorides of cobalt and
              manganese",
  author   = "Borovik-Romanov, A S",
  journal  = "Sov. Phys., JETP",
  volume   =  11,
  number   =  4,
  pages    = "786--786",
  year     =  1960
}

@ARTICLE{Baruchel1988-pl,
  title     = "{PIEZOMAGNETISM} {AND} {DOMAINS} {IN} {MnF2}",
  author    = "Baruchel, J and Draperi, A and El Kadiri, M and Fillion, G and
               Maeder, M and Molho, P and Porteseil, J L",
  journal   = "J. Phys., Colloq.",
  publisher = "EDP Sciences",
  volume    =  49,
  number    = "C8",
  pages     = "C8--1895--C8--1896",
  month     =  dec,
  year      =  1988
}

@ARTICLE{Moriya1959-gy,
  title     = "Piezomagnetism in {CoF2}",
  author    = "Moriya, T",
  journal   = "J. Phys. Chem. Solids",
  publisher = "Elsevier BV",
  volume    =  11,
  number    = "1-2",
  pages     = "73--77",
  month     =  "1~" # sep,
  year      =  1959
}

@ARTICLE{Dzialoshinskii1958-xq,
  title    = "The Problem of Piezomagnetism",
  author   = "Dzialoshinskii, I E",
  journal  = "Sov. Phys. JETP",
  volume   =  6,
  pages    = "621-622",
  year     =  1958
}

@ARTICLE{Dzialoshinskii1957-vg,
  title    = "Thermodynamic theory of `` weak '' ferromagnetism in
              antiferromagnetic substances",
  author   = "Dzialoshinskii, I E",
  journal  = "Sov. Phys. JETP",
  volume   =  5,
  number   =  6,
  pages    = "1259--1259",
  year     =  1957
}

@ARTICLE{Moriya1960-pu,
  title     = "New mechanism of anisotropic superexchange interaction",
  author    = "Moriya, Tôru",
  journal   = "Phys. Rev. Lett.",
  publisher = "American Physical Society (APS)",
  volume    =  4,
  number    =  5,
  pages     = "228--230",
  month     =  "1~" # mar,
  year      =  1960
}

@article{AoyamaPRM2024,
  author  = {Aoyama, Takuya and Ohgushi, Kenya},
  title   = {Piezomagnetic properties in altermagnetic MnTe},
  journal = {Phys. Rev. Mater.},
  volume  = {8},
  pages   = {024408},
  year    = {2024},
  doi     = {10.1103/PhysRevMaterials.8.024408}
}

@ARTICLE{Nanjo2025-lk,
  title         = "Piezomagnetic effect in 5$d$ transition metal oxides
                   {Y}$_$\_{2$}${Ir}$_$\_{2$}${O}$_$\_{7$}$ and
                   {Cd}$_$\_{2$}${Os}$_$\_{2$}${O}$_$\_{7$}$ with all-in/all-out
                   magnetic order",
  author        = "Nanjo, Hiroki and Imai, Yoshinori and Aoyama, Takuya and
                   Yamaura, Junichi and Ohgushi, Kenya",
  journal       = "arXiv [cond-mat.str-el]",
  month         =  "16~" # may,
  year          =  2025,
  archivePrefix = "arXiv",
  primaryClass  = "cond-mat.str-el"
}

@ARTICLE{Takagi2025-sl,
  title     = "Spontaneous Hall effect induced by collinear antiferromagnetic
               order at room temperature",
  author    = "Takagi, Rina and Hirakida, Ryosuke and Settai, Yuki and Oiwa,
               Rikuto and Takagi, Hirotaka and Kitaori, Aki and Yamauchi, Kensei
               and Inoue, Hiroki and Yamaura, Jun-Ichi and Nishio-Hamane,
               Daisuke and Itoh, Shinichi and Aji, Seno and Saito, Hiraku and
               Nakajima, Taro and Nomoto, Takuya and Arita, Ryotaro and Seki,
               Shinichiro",
  journal   = "Nat. Mater.",
  publisher = "Springer Science and Business Media LLC",
  volume    =  24,
  number    =  1,
  pages     = "63--68",
  month     =  jan,
  year      =  2025
}

@ARTICLE{Naka2019-px,
  title     = "Spin current generation in organic antiferromagnets",
  author    = "Naka, Makoto and Hayami, Satoru and Kusunose, Hiroaki and Yanagi,
               Yuki and Motome, Yukitoshi and Seo, Hitoshi",
  journal   = "Nat. Commun.",
  publisher = "Springer Science and Business Media LLC",
  volume    =  10,
  number    =  1,
  pages     =  4305,
  month     =  "20~" # sep,
  year      =  2019
}

@ARTICLE{Okugawa2018-is,
  title     = "Weakly spin-dependent band structures of antiferromagnetic
               perovskite {LaMO3} ({M} = Cr, Mn, Fe)",
  author    = "Okugawa, Takuya and Ohno, Kaoru and Noda, Yusuke and Nakamura,
               Shinichiro",
  journal   = "J. Phys. Condens. Matter",
  publisher = "IOP Publishing",
  volume    =  30,
  number    =  7,
  pages     =  075502,
  month     =  "21~" # feb,
  year      =  2018
}

@ARTICLE{Noda2016-wh,
  title     = "Momentum-dependent band spin splitting in semiconducting {MnO2}:
               a density functional calculation",
  author    = "Noda, Yusuke and Ohno, Kaoru and Nakamura, Shinichiro",
  journal   = "Phys. Chem. Chem. Phys.",
  publisher = "Royal Society of Chemistry (RSC)",
  volume    =  18,
  number    =  19,
  pages     = "13294--13303",
  month     =  "11~" # may,
  year      =  2016
}

@ARTICLE{Streltsov2025-ev,
  title     = "Altermagnetism in {6H} perovskites",
  author    = "Streltsov, Sergey V and Cheong, Sang-Wook",
  journal   = "Npj Quantum Mater.",
  publisher = "Springer Science and Business Media LLC",
  volume    =  10,
  number    =  1,
  pages     =  102,
  month     =  "31~" # oct,
  year      =  2025
}

@ARTICLE{Jungwirth2026-pd,
  title     = "Altermagnetic spintronics",
  author    = "Jungwirth, T and Sinova, J and Wadley, P and Kriegner, D and
               Reichlová, H and Krizek, F and Ohno, H and Šmejkal, L",
  journal   = "Nat. Phys.",
  publisher = "Springer Science and Business Media LLC",
  volume    =  22,
  number    =  7,
  pages     = "1012--1021",
  month     =  "6~" # jul,
  year      =  2026
}

@ARTICLE{Jungwirth2026-rn,
  title     = "Symmetry, microscopy and spectroscopy signatures of
               altermagnetism",
  author    = "Jungwirth, Tomas and Sinova, Jairo and Fernandes, Rafael M and
               Liu, Qihang and Watanabe, Hikaru and Murakami, Shuichi and
               Nakatsuji, Satoru and Šmejkal, Libor",
  journal   = "Nature",
  publisher = "Springer Science and Business Media LLC",
  volume    =  649,
  number    =  8098,
  pages     = "837--847",
  month     =  "21~" # jan,
  year      =  2026
}

@article{Momma2011,
author = {Momma, Koichi and Izumi, Fujio},
title = {VESTA3 for three-dimensional visualization of crystal, volumetric and morphology data},
journal = {J. Appl. Crystallogr.},
volume = {44},
number = {6},
pages = {1272-1276},
doi = {https://doi.org/10.1107/S0021889811038970},
url = {https://onlinelibrary.wiley.com/doi/abs/10.1107/S0021889811038970},
year = {2011}
}

@article{Kresse1996,
  title = {Efficient iterative schemes for ab initio total-energy calculations using a plane-wave basis set},
  author = {Kresse, G. and Furthm\"uller, J.},
  journal = {Phys. Rev. B},
  volume = {54},
  issue = {16},
  pages = {11169--11186},
  numpages = {0},
  year = {1996},
  month = {Oct},
  publisher = {American Physical Society},
  doi = {10.1103/PhysRevB.54.11169},
  url = {https://link.aps.org/doi/10.1103/PhysRevB.54.11169}
}

@article{Kresse1999,
  title = {From ultrasoft pseudopotentials to the projector augmented-wave method},
  author = {Kresse, G. and Joubert, D.},
  journal = {Phys. Rev. B},
  volume = {59},
  issue = {3},
  pages = {1758--1775},
  numpages = {0},
  year = {1999},
  month = {Jan},
  publisher = {American Physical Society},
  doi = {10.1103/PhysRevB.59.1758},
  url = {https://link.aps.org/doi/10.1103/PhysRevB.59.1758}
}

@article{Perdew1996,
  title = {Generalized Gradient Approximation Made Simple},
  author = {Perdew, John P. and Burke, Kieron and Ernzerhof, Matthias},
  journal = {Phys. Rev. Lett.},
  volume = {77},
  issue = {18},
  pages = {3865--3868},
  numpages = {0},
  year = {1996},
  month = {Oct},
  publisher = {American Physical Society},
  doi = {10.1103/PhysRevLett.77.3865},
  url = {https://link.aps.org/doi/10.1103/PhysRevLett.77.3865}
}

@article{Dudarev1998,
  author = {Dudarev, S. L. and Botton, G. A. and Savrasov, S. Y. and Humphreys, C. J. and Sutton, A. P.},
  title = {Electron-energy-loss spectra and the structural stability of nickel oxide: An LSDA+U study},
  journal = {Phys. Rev. B},
  volume = {57},
  pages = {1505--1509},
  year = {1998}
}

@article{Osumi2026RuO2,
  title = {Spin-degenerate bulk bands and topological surface states associated with Dirac nodal lines in ${\mathrm{RuO}}_{2}$},
  author = {Osumi, Takumi and Yamauchi, Kunihiko and Souma, Seigo and Paul, Shubhankar and Honma, Asuka and Nakayama, Kosuke and Ozawa, Kenichi and Kitamura, Miho and Horiba, Koji and Kumigashira, Hiroshi and Bigi, Chiara and Bertran, Fran\ifmmode \mbox{\c{c}}\else \c{c}\fi{}ois and Oguchi, Tamio and Takahashi, Takashi and Maeno, Yoshiteru and Sato, Takafumi},
  journal = {Phys. Rev. B},
  volume = {113},
  issue = {8},
  pages = {085116},
  numpages = {18},
  year = {2026},
  month = {Feb},
  publisher = {American Physical Society},
  doi = {10.1103/wvs6-hqfv},
  url = {https://link.aps.org/doi/10.1103/wvs6-hqfv}
}

@article{Osumi2024MnTe,
  title = {Observation of a giant band splitting in altermagnetic MnTe},
  author = {Osumi, T. and Souma, S. and Aoyama, T. and Yamauchi, K. and Honma, A. and Nakayama, K. and Takahashi, T. and Ohgushi, K. and Sato, T.},
  journal = {Phys. Rev. B},
  volume = {109},
  issue = {11},
  pages = {115102},
  numpages = {8},
  year = {2024},
  month = {Mar},
  publisher = {American Physical Society},
  doi = {10.1103/PhysRevB.109.115102},
  url = {https://link.aps.org/doi/10.1103/PhysRevB.109.115102}
}

@article{Thao2023CaCrO3,
  title = {Ab initio prediction of anomalous Hall effect in antiferromagnetic ${\mathrm{CaCrO}}_{3}$},
  author = {Nguyen, Thi Phuong Thao and Yamauchi, Kunihiko},
  journal = {Phys. Rev. B},
  volume = {107},
  issue = {15},
  pages = {155126},
  numpages = {9},
  year = {2023},
  month = {Apr},
  publisher = {American Physical Society},
  doi = {10.1103/PhysRevB.107.155126},
  url = {https://link.aps.org/doi/10.1103/PhysRevB.107.155126}
}

\section*{Supplementary Materials}
\section{Structural changes under \texorpdfstring{$xy$}{xy} shear strain}

\begin{figure}[tb]
\begin{center}
\includegraphics[width=0.45\textwidth]{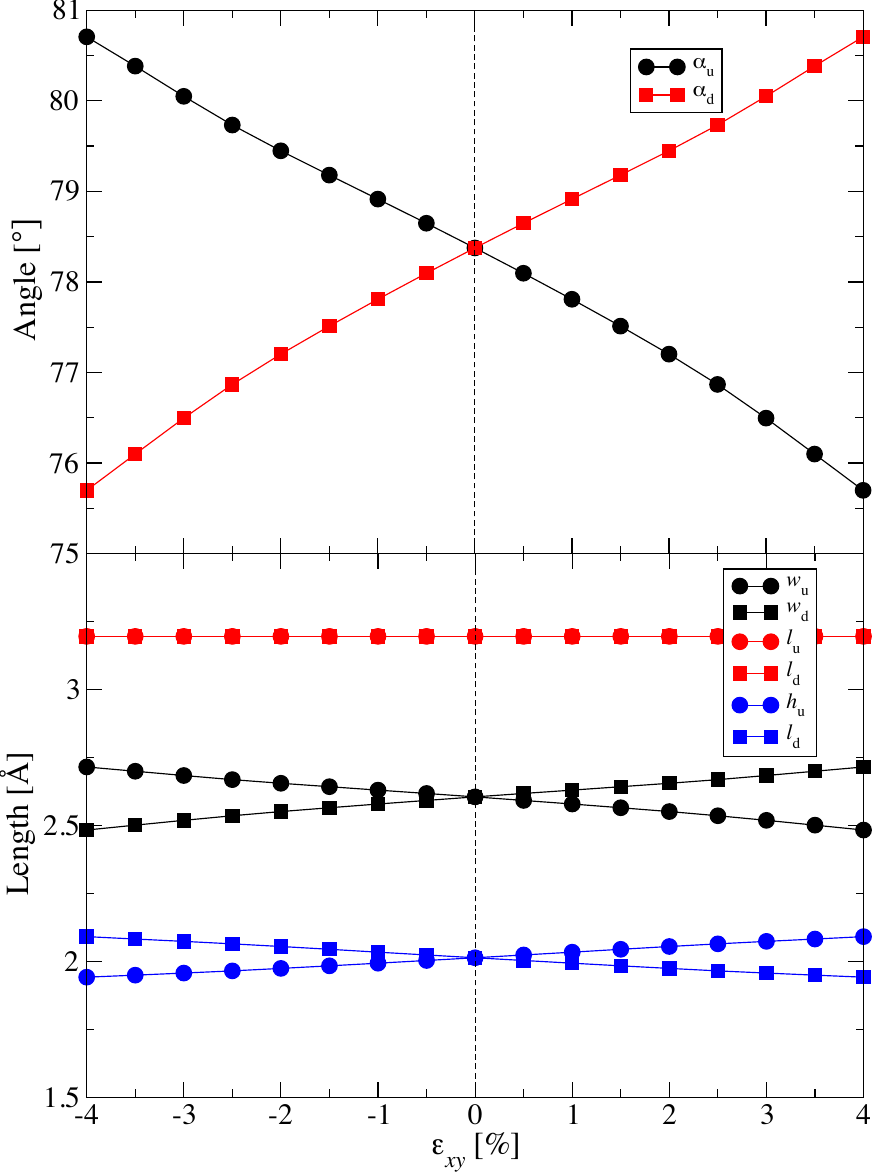}
\caption{
Structural changes of the CoF$_6$ octahedra under the $xy$ shear strain; see Fig.~1(a) of the main text for the definition of the structural parameters.
The upper panel shows the F--Co--F bond angle for Co1 (black circles) and Co2 (red squares).
The lower panel shows the equatorial bond lengths and the apical bond length of the CoF$_6$ octahedra around Co1 and Co2.
}
\label{fig:length_angle_suppl}
\end{center}
\end{figure}

The structural parameters shown in Fig.~\ref{fig:length_angle_suppl} provide additional information on the strain-induced distortion of the CoF$_6$ octahedra discussed in the main text.

\begin{figure*}[tb]
\begin{center}
    \includegraphics[width=0.95\textwidth]{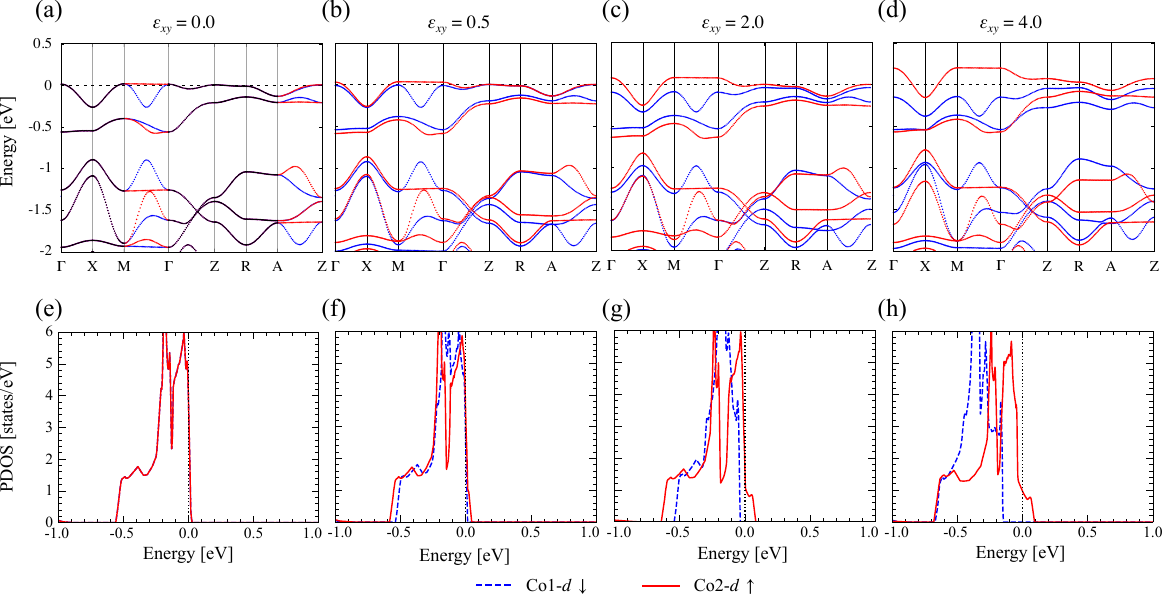}
    \caption{Band structures and partial densities of states of the minority-spin $d$ orbitals of Co1 and Co2 for 0.1\% hole-doped CoF$_2$ under $xy$ shear strains of $\varepsilon_{xy}=0.0$, 0.5, 2.0, and 4.0. Panels (a)–(d) show the band structures, while (e)–(h) show the corresponding partial densities of states. In (a)–(d), the red and blue dots denote the spin-up and spin-down bands, respectively, and the spin-degenerate bands at zero strain are shown by black lines. In (e)–(h), the minority-spin states of Co1 and Co2 correspond to the down-spin and up-spin channels, respectively. The energy is measured from the Fermi level.}
    \label{fig:bands_PDOS_strain-xy}
\end{center}
\end{figure*}

Figure~\ref{fig:bands_PDOS_strain-xy} provides the electronic origin of the nonlinear increase and subsequent saturation of the induced magnetization under the $xy$ shear strain. As shown in Figs.~\ref{fig:bands_PDOS_strain-xy}(a)--(d), the relatively flat valence-band maximum (VBM) along the M--$\Gamma$ line exhibits an additional spin splitting under the $xy$ shear strain. In the hole-doped case, the split bands move across the Fermi level around $\varepsilon_{xy}\sim2\%$, where the upper branch shifts above $E_{\rm F}$ and the lower branch moves below $E_{\rm F}$. This change in the band occupation is accompanied by a redistribution of electrons between the two inequivalent Co sites, leading to different local magnetic moments and a rapid increase in the induced magnetization. At larger strains, the occupation redistribution becomes nearly complete, and the induced magnetization consequently approaches saturation.

The same behavior is clearly reflected in the spin-resolved partial densities of states shown in Figs.~\ref{fig:bands_PDOS_strain-xy}(e)--(h). The minority-spin $d$ states of Co1 and Co2 evolve asymmetrically with increasing strain, directly visualizing the redistribution of the electronic occupation between the two Co sites. These results indicate that the nonrelativistic piezomagnetic response under the $xy$ shear strain originates from strain-induced electron transfer between the inequivalent Co sites, which changes the magnitudes of the local magnetic moments.

\begin{figure}[tb]
\begin{center}
\includegraphics[width=0.45\textwidth]{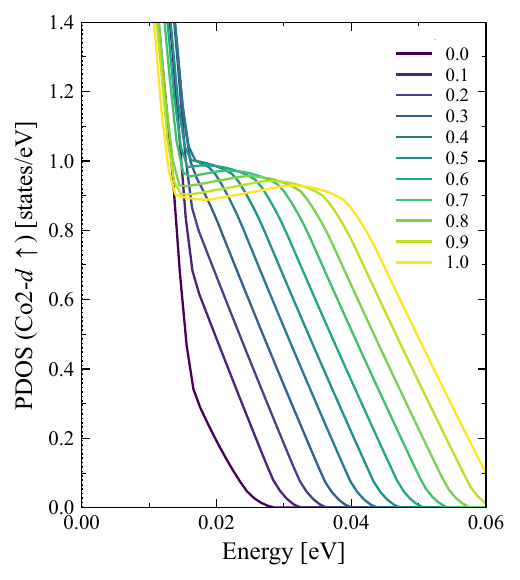}
\caption{
Evolution of the minority-spin $d$-orbital density of states of Co2 as a function of the $xy$ shear strain.
The legend indicates the applied strain $\varepsilon_{xy}$.
The energy is measured from the Fermi level.
}
\label{fig:Co2_d_up_DOS_suppl}
\end{center}
\end{figure}

Figure~\ref{fig:Co2_d_up_DOS_suppl} illustrates how the minority-spin density of states evolves under the $xy$ shear strain, complementing the discussion of the electronic mechanism presented in the main text.

\section{Stress-form piezomagnetic coefficients}

\begin{table}[htb]
\caption{
Calculated shear stresses, induced magnetic moments per unit cell, and unit-cell volumes for the $xy$ and $yz$ shear strains. Each unit cell contains two Co atoms. The stress signs follow the convention used in the VASP output.
}
\label{tab:stress}
\begin{center}
\begin{tabular}{ccccc}
\hline
Deformation & Strain & Shear stress & Magnetic moment & Volume \\
 & (\%) & (kbar) & ($\mu_{\rm B}$/cell) & (\AA$^3$) \\
\hline
$xy$, hole-doped, no SOC & 0.1 & $-1.369$  & $M_z=-0.007$ & 70.22 \\
$xy$, hole-doped, no SOC & 0.2 & $-2.774$  & $M_z=-0.015$ & 70.22 \\
$xy$, hole-doped, no SOC & 0.5 & $-10.353$ & $M_z=-0.094$ & 70.22 \\
\hline
$yz$, pristine, SOC & 0.5 & $-3.432$ & $M_x=0.010$ & 71.95 \\
$yz$, pristine, SOC & 1.0 & $-7.032$ & $M_x=0.019$ & 71.94 \\
\hline
\end{tabular}
\end{center}
\end{table}

Table~\ref{tab:stress} summarizes the calculated shear stresses, induced magnetic moments, and unit-cell volumes used to evaluate the piezomagnetic coefficients. The magnetic moments are given per unit cell containing two Co atoms, and the stress signs follow the VASP convention. The strain-form piezomagnetic coefficients were evaluated from the linear regions, namely $|\varepsilon_{xy}|\le0.5$ for the $xy$ shear strain and $|\varepsilon_{yz}|\le1.0$ for the $yz$ shear strain.

To convert the strain-form coefficients into the stress-form coefficients, the relaxed-ion elastic tensor $\bm C$ was calculated including spin--orbit coupling without imposing crystal symmetry because the noncollinear magnetic structure lowers the crystal symmetry. The compliance tensor was then obtained as
\[
\bm S=\bm C^{-1},
\]
and the resulting shear compliances are listed in Table~\ref{tab:compliance}.

\begin{table}[h]
\centering
\caption{Relaxed-ion shear compliances (TPa$^{-1}$).}
\label{tab:compliance}
\small
\begin{tabular}{lccc}
\hline\hline
State & $S_{44}$ (YZ) & $S_{55}$ (ZX) & $S_{66}$ (XY)\\
\hline
pristine & 25.03803 & 24.94304 & 11.76589\\
hole-doped & 23.83062 & 23.78146 & 13.06448\\
\hline\hline
\end{tabular}
\end{table}

The stress-form piezomagnetic tensor is related to the strain-form tensor by
\[
\Lambda_{ijk}=e^{\rm m}_{ilm}S_{lmjk}.
\]
The nonzero Cartesian components are
\begin{align}
\text{hole-doped:}\quad
&\Lambda_{123}=\Lambda_{132}
=\Lambda_{213}=\Lambda_{231}
=1.4\times10^{-12},\nonumber\\
&\Lambda_{312}=\Lambda_{321}
=-4.2\times10^{-12};\\
\text{pristine:}\quad
&\Lambda_{123}=\Lambda_{132}
=\Lambda_{213}=\Lambda_{231}
=1.8\times10^{-12},\nonumber\\
&\Lambda_{312}=\Lambda_{321}=0.0\times10^{-12}.
\end{align}

The corresponding coefficients in Voigt notation are summarized in Table~\ref{tab:results}.

\begin{table}[h]
\centering
\caption{Calculated stress-form piezomagnetic coefficients in Voigt notation (Wb/N).}
\label{tab:results}
\small
\begin{tabular}{lcc}
\hline\hline
State & $\Lambda_{14}(=\Lambda_{25})$ & $\Lambda_{36}$\\
\hline
pristine & $3.7\times10^{-12}$ & $0.0\times10^{-12}$\\
hole-doped & $2.8\times10^{-12}$ & $-8.4\times10^{-12}$\\
\hline\hline
\end{tabular}
\end{table}

For comparison, Borovik-Romanov \textit{et al.} reported $\Lambda_{14}=2.7\times10^{-11}$ and $\Lambda_{36}=1.0\times10^{-11}$ Wb/N for a single crystal of CoF$_2$ at 20.4 K.\cite{Borovik-Romanov1960-mi} 
Komuro \textit{et al.}\ reported an average piezomagnetic coefficient of $8.3\times10^{-2}$ emu/mol/MPa for a powder sample at 20 K,\cite{Komuro2025-ot} which corresponds to $\Lambda^{\rm Ave}=4.8\times10^{-12}$ Wb/N, where
\[
\Lambda^{\rm Ave}=\frac{|2\Lambda_{14}+\Lambda_{36}|}{2\pi}.
\]
The corresponding averaged coefficients obtained from the present calculations are $1.2\times10^{-12}$ and $4.5\times10^{-13}$ Wb/N for pristine and hole-doped CoF$_2$, respectively. The calculated values are within approximately one order of magnitude of the experimental value. 
When comparing with experiment, it should be noted that the sign convention for the calculated magnetization follows the spin direction adopted in the present calculations. Therefore, care should be taken when comparing the signs of the calculated and experimental piezomagnetic coefficients. Given that the available single-crystal data date back to early measurements, updated experimental determination of the piezomagnetic coefficients would be valuable for a more quantitative comparison with the present calculations.

\end{document}